\documentclass[a4paper]{jpconf}
\usepackage{graphicx}
\usepackage{amssymb}
\begin{document}
%
\def\d{{\mathrm{d}}}
\def\r{{\mathbf{r}}}
\newcommand{\norm}[1]{\left\Vert#1\right\Vert}
\newcommand{\betrag}[1]{\left\vert#1\right\vert}
\newcommand{\bra}[1]{\langle #1\vert}
\newcommand{\ket}[1]{\arrowvert #1 \rangle}
\newcommand{\braket}[1]{\langle #1\rangle}
\newcommand{\kin}{-\frac{\hbar^2}{2m} \, \nabla^2 \,}
\newcommand{\oppdag}[2]{\hat{#1}^\dagger(t,\vec{#2})}
\newcommand{\opp}[2]{\hat{#1}(t,\vec{#2})}
\newcommand{\bog}{\hat{\Psi}(t,\vec{x})=\Phi(t,\vec{x})
  \, + \, \varepsilon \hat{\psi}(t,\vec{x})\, + \,
  \ldots} \newcommand{\zeit}{i \,\hbar \,
  \frac{\partial}{\partial t} \,}
\newcommand{\mad}{\sqrt{\rho(t,\vec{x})}e^{i\theta(t,\vec{x})}e^{-\frac{i\mu
      t}{\hbar}}}
\def\ie{{\em i.e.\/}}
 \def\eg{{\em e.g.\/}}
\def\etc{{\em etc.\/}}
 \def\etal{{\em et al.\/}}
\def\half{{1\over2}} 
\def\L{{\mathcal L}}
\def\d{{\mathrm{d}}} 
\def\x{{\mathbf x}} 
\def\v{{\mathbf v}} 
\def\im{{\rm i}}
\def\etal{{\emph{et al\/}}} 
\def\det{{\mathrm{det}}}
\def\tr{{\mathrm{tr}}} 
\def\ie{{\emph{i.e.}}}
\def\bnabla{\mbox{\boldmath$\nabla$}}
\def\Box{\kern0.5pt{\lower0.1pt\vbox{\hrule height.5pt
      width 6.8pt \hbox{\vrule width.5pt height6pt
        \kern6pt \vrule width.3pt} \hrule height.3pt
      width 6.8pt} }\kern1.5pt}
\def\HRULE{{\bigskip\hrule\bigskip}}
\def\Schrodinger{Schr\"odinger}

\title{Modelling  Planck-scale Lorentz violation via analogue models}
\author{Silke Weinfurtner$^1$, Stefano Liberati$^2$, and Matt Visser$^1$}

\address{$^1$
School of Mathematics, Statistics,  and Computer Science,\\
Victoria University of Wellington, PO Box 600, Wellington, New Zealand}

\address{$^2$ International School for Advanced Studies, \\
Via Beirut 2-4, 34013 Trieste, Italy, and INFN, Trieste}
\ead{silke.weinfurtner@mcs.vuw.ac.nz, liberati@sissa.it, matt.visser@mcs.vuw.ac.nz }
\date{24 November 2005; \LaTeX-ed \today}
\begin{abstract}
Astrophysical tests of Planck-suppressed Lorentz violations had been extensively studied in recent years and very stringent constraints have been obtained within the framework of effective field theory. There are however still some unresolved theoretical issues, in particular regarding the so called ``naturalness problem'' --- which arises when postulating that Planck-suppressed Lorentz violations arise only from operators with mass dimension greater than four in the Lagrangian.
In the work presented here we shall try to address this problem by looking at a condensed-matter analogue of the Lorentz violations considered in quantum gravity phenomenology.   Specifically, we investigate the class of two-component BECs subject to laser-induced transitions between the two components,  and we show that this model is an example for Lorentz invariance violation due to ultraviolet physics.     
We shall show that such a model can be considered to be an explicit example high-energy Lorentz violations where the ``naturalness problem'' does not arise.
\end{abstract}

\section{Introduction and motivation}

The search for a quantum theory encompassing gravity has been a major issue in theoretical
physics for the last 60 years. Nonetheless, until recently quantum gravity was largely relegated to the
realm of speculation due to the complete lack of observational or
experimental tests. In fact the traditional scale of
quantum gravitational effects, the Planck scale $M_{\rm
  Planck}=1.2\times10^{19}$ GeV/$c^2$, is completely out of reach for any
experiment or observation currently at hand. This state of affairs had
led the scientific community to adopt the 
``folklore'' that testing quantum
gravity was completely impossible.
However the last decade has seen a dramatic change in
this respect, and nowadays one can encounter a growing
literature dealing with tests of possible predictions of various
quantum gravity models~\cite{LIV, Jacobson:2002hd, TeV-QG, others}. This
field goes generically under the name of ``quantum
gravity phenomenology".

Among the several generic predictions associated with quantum
gravity models, the possibility that Planck-scale physics might induce
violations of Lorentz invariance has played a particularly important role~\cite{LIV, Jacobson:2002hd}. 
Generically any possible ``discreteness'' or ``granularity'' of spacetime at the Planck scale seems
incompatible with strict Lorentz invariance (although some
particular quantum gravity theories might still preserve it, see
e.g.~\cite{Rovelli}) since larger and larger boosts expose shorter and shorter distances. 
Actually we now have a wealth of theoretical studies --- for example in the context of string field theory~\cite{KS89, Damour:1994zq}, spacetime foam scenarios~\cite{GAC-Nat},
semiclassical calculations in loop quantum gravity~\cite{GP,loopqg}, deformed special relativity models~\cite{DSR, DSRint, Judes}, or
non-commutative
geometry~\cite{Hayakawa,Mocioiu:2000ip,Carroll:2001ws,Anisimov:2001zc}, just to cite a few --- all leading to high-energy violations of Lorentz invariance.

Interestingly most investigations, even if they arise from quite different 
fundamental physics, seem to converge on the prediction that the breakdown of Lorentz invariance generically becomes manifest in the form of modified dispersion relations exhibiting extra energy-dependent  or momentum-dependent terms, apart from the usual quadratic one occurring  in the Lorentz
invariant dispersion relation:
%
$E^2=m^2\;c^4+p^2\;c^2.$
%
In the absence of a definitive theory of quantum gravity it became
common to adopt,  in most of the literature seeking to put these predictions to observational test, a purely
phenomenological approach, \ie,~one that modifies the dispersion relation by adding some generic momentum-dependent (or
energy-dependent) function $F(p,c,M_{\rm Planck})$ to be expanded in powers of the dimensionless quantity $p/(M_{\rm Planck}\; c)$. Hence the ansatz reads:
\begin{eqnarray}
E^2&=&
m^2\;c^4+p^2\;c^2+M_{\rm Planck}^2\; c^4\left\{\sum_{n\geq1}
\eta_n\,\frac{p^{n}}{(M_{\rm Planck} \; c)^n}\right\};
\\
&=&
m^2\;c^4+p^2\;c^2+c^4\left\{\eta_1\, M_{\rm Planck}\, p/c+\eta_2\,p^2/c^2+\sum_{n\geq3}
\eta_n\,\frac{(p/c)^{n}}{M^{n-2}_{\rm Planck}}\right\};
\label{eq:mod-disp}
\end{eqnarray}
where  the $\eta_n$ are chosen to be dimensionless. Since these dispersion relations are not Lorentz
invariant, it is necessary to specify the particular inertial frame in which they are given, and generally one chooses the cosmic microwave background (CMB) frame. 
Of course compatibility with low-energy experiments implies that somehow the $\eta_1$ and $\eta_2$ coefficients must be extremely small (and indeed very stringent constraints on these two terms have been obtained, \emph{e.g.}, within the framework of the so called Standard Model Extension~\cite{SME}). Hence most of the attention in the recent years have been focussed on the higher-order (explicitly Planck suppressed) $n\geq 3$ Lorentz violating terms in  Eq~(\ref{eq:mod-disp}). 

At first sight it might appear hopeless to constrain such deviations from exact Lorentz invariance without exploring Planckian regimes. However a more careful analysis shows that no ultra high energies are needed for such a task (at least for powers with $n=3,4$). The basic idea is that there are some special phenomena where the tiny violations like those present in Eq~(\ref{eq:mod-disp}) can be amplified by suitable physical effects.  Let us give a simple example by considering a dispersion relation like~(\ref{eq:mod-disp}) but with only one Lorentz-violating term suppressed by some power of $p/(M_{\rm Planck}\; c)$:
\begin{equation}
E^2=c^4\left\{m^2+p^2/c^2+\eta_n\,\frac{(p/c)^{n}}{M^{n-2}_{\rm Planck}}\right\}.
\label{eq:ex-disp}
\end{equation}
If we now consider a threshold reaction involving the creation of such a particle it is easy to see that it will \emph{not} be necessary to require the Lorentz-violating term to be comparable to the $p^2$ term in order to have a detectable effect. Indeed the threshold scale will be fixed by the mass scale $m$ and the modification should be expected to become important when the first and last term on the l.h.s.~of Eq~(\ref{eq:ex-disp}) are comparable.
It is east to see that, assuming $\eta_n=O(1)$, this happens when
\begin{equation}
p/c\approx \left[m^2\;M_{\rm Planck}^{n-2}\right]^{1/n}
\end{equation}
which indeed, for $n=3,4$, corresponds to energy scales not totally beyond reach in high energy astrophysics. For example the above equation for $n=3$ implies that one can cast a constraint of order one on the dimensionless coefficient $\eta_3$ for, say, an electron, by looking at a threshold reaction involving such particles with energies of about 10 TeV.

Indeed threshold reactions are not the only phenomena accessible to current observations/\-experiments which are sensitive to possible violations of
Lorentz invariance. A partial list is:
\begin{itemize}
\item dynamical effects of LV background fields (\emph{e.g.}~gravitational coupling and additional gravitational wave modes);
\item sidereal variation of LV couplings as the lab moves
  with respect to a preferred frame or directions, or cosmological variation;
\item long baseline dispersion and vacuum birefringence
  (\emph{e.g.}~of signals from gamma ray bursts, active galactic
  nuclei, pulsars, galaxies);
\item new reaction thresholds (\emph{e.g.}~photon decay, vacuum
Cherenkov effect);
\item shifted thresholds (\emph{e.g.}~photon annihilation from
  blazars or GZK reaction \emph{i.e.}~pion production by collision of ultra high energy protons with CMB photons);
\item maximum velocity (\emph{e.g.}~synchrotron peak from supernova
remnants).
\end{itemize}

\section{EFT with high energy Lorentz violations: a brief overview}

Of course merely specifying a set of dispersion relations is not always enough to place significant constraints on the model --- as most observations need at least some assumption on the dynamics for their interpretation. In fact most of the available constraints are extracted from some assumed model ``test theory". 
Although several alternative scenarios have been considered in the literature, 
so far the most commonly explored avenue is an effective field theory (EFT) approach (see \eg,~\cite{jlm-ann} for a review focussed on this framework, and ~\cite{jlm-notes, jlm-qgp, jlm-limits, jlm-comments, jlm-nature,  jlm-tev, jlm-bloomington} for some of the primary literature). The main reasons for this choice can be summarized in the fact that we are very familiar with this class of theories, and that it is a widely accepted idea (although not unanimously accepted, see \eg~\cite{GAC-crit}) that \emph{any} quantum gravity scenario should admit a suitable EFT description at low energies.  All in all, the standard model and general relativity itself, (which are presumably not fundamental theories) are EFTs, as are most models of condensed-matter systems at appropriate length and energy scales. Even ``fundamental'' quantum gravity candidates such as string theory admit an EFT description at low energies (as perhaps most impressively verified in the calculations of black hole entropy and
Hawking radiation rates). 

The EFT approach to the study of Lorentz violations has been remarkably useful in the last few years. Nowadays the best studied theories incorporating Lorentz violations are EFTs where Lorentz violations  are associated  either with renormalizable Lorentz-violating operators (mass dimension four or less), or sometimes with higher-order Lorentz-violating operators (mass dimension 5 or greater, corresponding to order $p$ and $p^2$ deviations in the dispersion relation~(\ref{eq:mod-disp})). The first approach is generally known as the Standard Model Extension~\cite{SME}, while the second has been formalized by Myers and Pospelov~\cite{MP} in the form of QED with dimension five Lorentz-violating operators (order $p^3$ deviations in the dispersion relation of Eq~(\ref{eq:mod-disp}). In both cases extremely accurate constraints have been obtained using a combination of  experiments as well as observation (mainly in high energy astrophysics). 

Let us focus for the moment on the second case, \emph{i.e.} on an ansatz in which one presumes  that the coefficients $\eta_1$ and $\eta_2$ are absent or extremely small (as Lorentz invariance is very well tested at low energies), while the first relevant Lorentz-violating term is associated with a cubic power of the momentum. We shall first briefly review the main constraints obtained so far in this framework and later discuss some open issues which we shall finally try to address with a condensed matter analogue model.

\section{Observational constraints on $O(E/M)$ Lorentz violations in QED}

As discussed above,  the study of Lorentz-violating EFT in the higher mass dimension sector was
initiated by Myers and Pospelov~\cite{MP}.  They classified all LV
dimension-five operators that can be added to the QED Lagrangian
and are quadratic in the same fields, rotation invariant, gauge
invariant, not reducible to a combination of lower and/or higher
dimension operators using the field equations, and contribute
$p^3$ terms to the dispersion relation. 

In the limit of high energy $E\gg m$, the photon  and electron
dispersion relations following from QED with the above terms
are~\cite{MP}:
\begin{eqnarray}
\omega_{\pm}^2&=& k^2 \pm \frac{\xi}{M_{\rm Planck}}k^3\\
E_{\pm}^2&=& p^2 + m^2  +\frac{2(\zeta_1\pm\zeta_2)}{M_{\rm Planck}}p^3,
\label{QEDdisp} \end{eqnarray}
where $\xi,\zeta_{1,2}$ are dimensionless parameters (and $\hbar=c=1$).
The photon subscripts $\pm$ refer to helicity, \emph{i.e.} right and left
circular polarization,
which it turns out necessarily have opposite LV parameters. The
electron subscripts $\pm$ also refer to the helicity, which can be
shown to be a good quantum number in the presence of these LV
terms~\cite{jlm-limits}.
Moreover, if we write
$\eta_\pm=2(\zeta_1\pm\zeta_2)$ for the LV parameters of the two
electron helicities, those for positrons
are given~\cite{jlm-limits} by 
\begin{equation} 
\label{eq:ep} \eta^{\rm
positron}_\pm=-\eta^{\rm electron}_\mp. 
\end{equation}

\subsection{Constraints}

We now provide a partial and brief list of the current best constraints on extended QED with dimension 5 Lorentz violations. The main aim of such a partial overview is simply to provide us with quantitative insight about the current strength of the constraints and how they are obtained. For details, we direct the reader to the much more extensive accounts of these constraints currently available in the literature (see \emph{e.g.}~\cite{LIV} and \cite{jlm-ann}).
 
\paragraph{Photon time of flight:}
Photon time of flight constraints~\cite{TOF} limit differences
in the arrival time on Earth for photons originating in a distant
event~\cite{GAC-Nat}. Time of flight can vary with
energy since the LV term in the group velocity is $\xi k/M_{\rm Planck}$. The
arrival time difference for wave-vectors $k_1$ and $k_2$ is thus
\begin{equation} 
\Delta t=\xi (k_2-k_1) d/M_{\rm Planck},
  \label{timediff}
\end{equation}
which is proportional to the energy difference and the distance
travelled.  Constraints were obtained using the high energy radiation emitted by some gamma ray bursts (GRB) and active galaxies of the blazar class. The strength of such constraints is typically $\xi\lesssim {\cal O}(100)$ or weaker~\cite{TOF, EllisNew}.

\paragraph{Vacuum birefringence:}
Looking at Eq~(\ref{QEDdisp}) one sees that the parameters for the Lorentz violation for left and right circular polarized photons are equal and
opposite.  This implies that the phase velocity thus depends on
both the wavevector and the helicity. Linear polarization is
therefore rotated through an energy-dependent angle as a signal
propagates, which depolarizes an initially linearly polarized signal
comprised of a range of wavevectors. Hence the observation of
linearly polarized radiation coming from far away can constrain the
magnitude of the LV parameter. Unfortunately we do not have at the moment reliable observations of polarized photon fluxes at very high energy. It is however remarkable that  the most reliable constraint on the dimension five term was deduced by Gleiser and Kozameh~\cite{GK} using UV light from distant galaxies, and is given by $|\xi|\lesssim 2\times10^{-4}$.

\paragraph{Threshold effects:}
There are several threshold effects that can be influenced by the presence of additional, Lorentz violating, terms like those in Eq~(\ref{QEDdisp}). In general we can distinguish three kinds of threshold reactions: 
\begin{enumerate}
\item[1)] Threshold reactions which are normally forbidden, but are allowed if Lorentz invariance is broken.
\item[2)] Threshold reactions which are normally allowed, but are sufficiently high-energy that their thresholds are non-negligibly shifted in the presence of Planck-suppressed Lorentz violations. 
\item[3)] Finally there are reactions which become allowed due to Lorentz violations but do not strictly speaking have a threshold energy, though they do have a very suppressed reaction rate. These reactions can be said to have an ``effective threshold energy'' above which the rate is non-negligible.
\end{enumerate}
Regarding the extended QED of Eq~(\ref{QEDdisp}), the first category includes two reactions associated with the basic QED vertex, \emph{i.e.}~photon decay, $\gamma\to e^+ e^-$ and vacuum Cherekov $e^\pm\to e^\pm \gamma$. A general property of these reactions is that above threshold they have a very rapid rate so that the particle under consideration would effectively be unable to propagate.  The second category instead includes the very important astrophysical reaction of gamma pair creation, $\gamma \gamma_0\to e^+ e^-$, where a TeV-scale gamma ray annihilates with a (IR or CMB) background photon to create a electron-positron pair. Finally, the third category includes reactions like photon splitting, $\gamma\to n\gamma$,  helicity decay, $e^\pm\to e^\pm \gamma$, or electron pair creation, $e^\pm\to e^\pm\, e^\pm e^\mp$. (See \cite{jlm-ann} for a detailed discussion.)

Threshold analyses have been very effective tools in constraining the modified dispersion relations (\ref{QEDdisp}). In particular the best constraint up to date on $|\eta_{\pm}|\lesssim10^{-1}$ on the coefficients for the electron/positron comes from the detection of 50 TeV inverse Compton gamma rays from the Crab nebula (which implies absence of photon decay up to this energies).

\paragraph{Synchrotron radiation:}
The Crab nebula also provides another interesting mechanism for constraining the dispersion relations (\ref{QEDdisp}). We do in fact detect a characteristic synchrotron emission from the Crab nebula, which can be used to cast a strong constraint on at least one of the $\eta$ coefficients.  Indeed an electron or positron with a negative value of $\eta$ has a maximal group velocity less than the low-energy speed of light. This implies an upper bound on the relativistic gamma factor for such a lepton, which in turn implies~\cite{Crab} the presence of a maximal synchrotron frequency $\omega_c^{\rm max}$ that such a lepton will be able to produce, regardless of its energy, in a given magnetic field $B$~\cite{Crab}. Thus for at least one electron or positron helicity $\omega_c^{\rm max}$ must be greater than the maximum observed synchrotron emission
frequency $\omega_{\rm obs}$. This yields the constraint
\begin{equation}
\eta > - \frac{M_{\rm Planck}}{m}\left(\frac{0.34\, eB}{m\,\omega_{\rm
obs}}\right)^{3/2}. \label{eq:synchcon}
\end{equation}
The strongest constraint is obtained in the case of a system that
has the smallest $B/\omega_{\rm obs}$ ratio. This occurs exactly in the Crab
nebula, which emits synchrotron radiation up to 100 MeV and has a
magnetic field no larger than 0.6 mG in the emitting region. Thus one can infer that at least one of $\pm\eta_\pm$ must be greater than $-7\times10^{-8}$. (Actually, to obtain a useful constraint this analysis must be complemented with a study of the inverse Compton peak --- 
for further details, see~\cite{jlm-ann}.)

\section{Open issues for a Lorentz violating EFT}
 
 In spite of the remarkable success in constraining the  modified QED of Eq~(\ref{QEDdisp}) it is interesting to note that there are still significant open issues concerning its theoretical foundations. In particular, let us now focus on  two aspects of this approach that have spurred some debate among members of the quantum gravity phenomenology
community.

\paragraph{The naturalness problem:}

Looking at the dispersion relation (\ref{eq:mod-disp}) it might seem that the deviations linear and quadratic in $p$ are \emph{not} Planck suppressed, and hence are always dominant (and grossly incompatible with observations).
However one might hope that there will be some other characteristic QFT mass scale $\mu\ll M_{\rm Planck}$ (\ie, some particle physics mass scale) associated with the Lorentz symmetry breaking which might enter in the lowest order dimensionless coefficients
$\eta_{1,2}$, which will be then generically suppressed by appropriate ratios of this characteristic mass to the Planck mass. 
Following the observational leads one might then assume behaviours like $\eta_1\propto (\mu /M_{\rm Planck})^{\sigma+1}$, and $\eta_2\propto (\mu/M_{\rm Planck})^\sigma$ where $\sigma\geq 1$ is some positive power (often taken as one or two). Meanwhile no extra Planck suppression is assumed for the higher order $\eta_n$ coefficients, which are then naturally of order one. Note that such an ansatz assures that the Lorentz-violating term linear in the particle momentum in Eq~(\ref{eq:mod-disp}) is always subdominant with respect to the quadratic one, and that the Lorentz-violating term cubic in the momentum is the least suppressed of the higher-order ones.~\footnote{Of course this is only true provided there is no symmetry like CPT that automatically cancels all the terms in odd powers of the momentum. In that case the least suppressed Lorentz-violating term would be that quartic in the momentum.}  If this is the case one will have two distinct regimes: For low momenta $p/(M_{\rm Planck}\;c) \ll (\mu/M_{\rm Planck})^\sigma$ the lower-order (quadratic in the momentum) deviations in~(\ref{eq:mod-disp}) will dominate over the higher-order (cubic and higher) ones, while at high energies $p/(M_{\rm Planck}\;c) \gg (\mu/M_{\rm Planck})^\sigma$ the higher-order terms (cubic and above in the momentum) will be dominant.

The naturalness problem arises because such a line of reasoning does not seem to be well justified within an EFT framework. In fact we implicitly assumed that there are no extra Planck suppressions hidden in the dimensionless coefficients $\eta_n$ with $n\geq 3$. Indeed we cannot justify why \emph{only} the dimensionless coefficients of the $n\leq 2$ terms should be suppressed by powers of the small ratio $\mu/M_{\rm Planck}$.  Even worse, renormalization group arguments seem to imply that a similar mass ratio, $\mu/M_{\rm Planck}$ would implicitly be present in \emph{all} the dimensionless $n\geq3$ coefficients --- hence suppressing them even further, to the point of complete undetectability.  Furthermore it is easy to show~\cite{Collins} that, without some protecting symmetry, it is generic that radiative corrections due to particle interactions in an  EFT with only Lorentz violations of order $n\geq 3$ in (\ref{eq:mod-disp}) for the free particles, will generate $n=1$ and $n=2$ Lorentz violating terms in the dispersion relation, which will then be dominant.  

\paragraph{The universality issue:}
The second point is not so much a problem,
as an issue of debate as to the best strategy to adopt. In dealing with situations with multiple particles one has to choose between the case of universal
(particle-independent) Lorentz violating coefficients
$\eta_n$, or instead go for a more general ansatz and
allow for particle-dependent coefficients; hence
allowing different magnitudes of Lorentz symmetry
violation for different particles even when considering
the same order terms (same $n$) in the momentum
expansion. The two choices are equally represented in the
extant literature (see \emph{e.g.}~\cite{GAC-Pir} and \cite{Jacobson:2002hd} for the two alternative ans\"atze), but it would be interesting to
understand how generic this universality might be, and what sort of processes might induce non-universal  Lorentz violation for different particles.

\vskip 20 pt

To shed some light on these issues it would definitely be
useful to have something that can play the role of test-bed for some of the ideas related to the emergence and form of the modified dispersion relations of Eq~(\ref{eq:mod-disp}).  In this regard,  herein we will consider an analogue model of
spacetime, that is, a condensed-matter system which admits
excitations whose propagation mimics that of quantum
fields on a curved spacetime~\cite{AM}.  Indeed it is well known
that the discreteness at small scales of such systems
shows up exactly via modified dispersion relations of the
kind described by Eq \eref{eq:mod-disp}, and one may hope that the complete
control over the microscopic (trans-Planckian) physics in
these systems would help understanding the nature of the
issues discussed above.

Following this physical insight we investigated the energy-dependent behavior of sound waves in a 2-component Bose--Einstein condensate (BEC). Building on the existence of so-called ``analogue models'' (AM) for minimally coupled massless fields in curved spacetimes~\cite{Unruh, unexpected, ergosphere, Garay, BEC1, AM, broken}, we shall show how to extend the AM to include massive particles \cite{2BEC}. As in the standard ansatz for quantum gravity phenomenology, the dispersion relation (now written in terms of frequency and wavenumber) is modified by Lorentz-violating terms at the ``Lorentz violation scale''  $M_{\rm eff}$ --- which in the present context is the most suitable definition of an effective ``analogue Planck scale'' \cite{QGPAM}:
\begin{equation} \label{ext_disp_rel}
\omega^2 = \omega_0^2 + \left(1 + \eta_{2} \right) \, c^2 \; k^2 + \eta_{4} \, \left(\frac{\hbar}{M_{\mathrm{eff}}} \right)^2  \; k^4 + \dots \; .
\end{equation}
The determination, by the microphysics of the system, of the dimensionless coefficients $\eta_{2}$ and $\eta_{4}$ for both massive and mass-less quasi-particles will finally allow us to discuss the naturalness problem and universality issue in our 2-component AM.

\section{Sound waves in 2-component BECs}

The basis for our AM is an ultra-cold dilute atomic gas of $N$ bosons, which exist in two single-particle states $\vert 1 \rangle$ and $\vert 2 \rangle$.    For example, we consider two different hyperfine states, $\vert F=1,m_{F}=1 \rangle$ and $\vert F=2,m_{F}=2 \rangle$ of $^{87}Rb$ \cite{jenkins,trippenbach}.  They have different total angular momenta $F$ and therefore slightly different energies. That permits us, from a theoretical point of view, to keep $m_{1} \neq m_{2}$, even if they are very nearly equal (to about one part in $10^{16}$). At the assumed ultra-cold temperatures and low densities the atoms interact only via low-energy collisions, and the 2-body atomic potential can be replaced by a contact potential. That leaves us with with three atom-atom coupling constants, $U_{11}$, $U_{22}$, and $U_{12}$, for the interactions within and between the two hyperfine states. For our purposes it is essential to include an additional laser field, that drives transition between the two single-particle states.~\footnote{A more detailed description on how to set up an external field driving the required transitions can be found in \cite{Bloch}.} The rotating-frame Hamiltonian for our closed 2-component system is given by:~\footnote{In general, it is possible that the collisions drive coupling to other hyperfine states. Strictly speaking the system is not closed, but it is legitimate to neglect this effect~\cite{dressed}. }
\begin{eqnarray}
\hat H = \int \d \r \;
\Bigg\{  \sum_{i = 1,2}  \left(-\hat \Psi_{i}^\dag \frac{\hbar^2 \nabla^2}{2 m_{i}} \hat \Psi_{i}  
+ \hat \Psi_i^\dag V_{ext,i} (\r) \hat \Psi_i \right) 
\nonumber 
+ \frac{1}{2} \sum_ {i,j = 1,2} \left(
 U_{i j} \hat \Psi_i^\dag  \hat \Psi_j^\dag  \hat \Psi_i  \hat \Psi_j  
 + \lambda  \hat \Psi_i^\dag (\mathbf{\sigma}_{x})_{i j}  \hat \Psi_j
\right)
\Bigg\} \, ,
\end{eqnarray}
with the transition energy $\lambda = \hbar \, \omega_{\mathrm{Rabi}}$ containing the effective Rabi frequency between the two hyperfine states. Here $\hat \Psi_i(\r)$ and $\hat \Psi_i^{\dag}(\r)$ are the usual boson field annihilation and creation operators for a single-particle state at position $\r$, and $\mathbf{\sigma}_x$ is the usual Pauli matrix.
For temperatures at or below the critical BEC temperature, almost all atoms occupy the spatial modes $\Psi_1(\r)$ and $\Psi_2(\r)$. The mean-field description for these modes,
\begin{equation}  \label{2GPE} 
 i \, \hbar \, \partial_{t} \Psi_{i} = \left[
   -\frac{\hbar^2}{2\,m_{i}} \nabla^2 + V_{i}-\mu_{i} + U_{ii}
   \, \betrag{\Psi_{i}}^2 + U_{ij} \betrag{\Psi_{j}}^2
   \right] \Psi_{i} + \lambda \, \Psi_{j} \, , 
\end{equation}
are a pair of coupled Gross--Pitaevskii equations (GPE):  $(i,j)\rightarrow (1,2)$ or  $(i,j)\rightarrow (2,1)$.

In order to use the above 2-component BEC as an AM, we have to investigate small perturbations (sound waves) in the condensate cloud.~\footnote{The perturbations have to be small compared to the overall size of the condensate could, so that the system remains in equilibrium.}  The excitation spectrum is obtained by linearizing around some background densities $\rho_{i0}$ and phases $\theta_{i0}$, using:
\begin{equation}
\Psi_{i}= \sqrt{\rho_{i0}+ \varepsilon \, \rho_{i1} }\,
e^{i(\theta_{i0}+ \varepsilon \, \theta_{i1} )}
\quad\hbox{for}\quad i=1,2 \, .
\end{equation}
A tedious calculation \cite{2BEC,QGPAM} shows that it is convenient to introduce the following $2 \times 2$ matrices: An effective coupling matrix,
\begin{equation}
\hat{\Xi}=\Xi+\hat X, 
\end{equation}
where
\begin{equation}
\Xi \equiv [\Xi]_{ij} =\frac{1}{\hbar}
\tilde{U}_{ij} = \frac{1}{\hbar} \left( U_{ij}-(-1)^{i+j} \, {\lambda\sqrt{\rho_{10}\rho_{20}}\over2} {1\over {\rho_{i0}\rho_{j0}} } \right)
\end{equation} 
and
\begin{equation}
\hat X \equiv  [\hat X]_{ij}= -{\hbar\over2} \, \delta_{ij} \, \frac{ \hat Q_{i1}}{m_i} 
= -{\hbar\over4}  \frac{ \delta_{ij} }{ m_i\;\rho_{i0}} 
= - [X]_{ij} \; \nabla^2 \, .
\end{equation}
Without transitions between the two hyperfine states, when $\lambda=0$, the matrix $\Xi$ only contains the coupling constants $[\Xi]_{ij} \rightarrow U_{ij}/\hbar$. While $\Xi$ is then independent of the energy of the perturbations, the importance of $\hat X$ increases with the energy of the perturbation. In the so-called hydrodynamic approximation, $\hat X$ can be neglected, $\hat X \rightarrow 0$, and $\hat \Xi \rightarrow \Xi$. 

Besides the interaction matrix, we also introduce a transition matrix,
\begin{equation}
\Lambda\equiv [\Lambda]_{ij}= -\frac{2\lambda\;\sqrt{\rho_{i0}\,\rho_{j0}} }{\hbar} \, (-1)^{i+j} \, ,
\end{equation}
and a mass-density matrix,
\begin{equation}
D\equiv [D]_{ij} = \hbar \, \delta_{ij} \,  \frac{ \rho_{i0} } {m_{i} } \, .
\end{equation} 

The final step is to define two column vectors,  ${\bar{\theta}} = [\theta_{11},\theta_{21}]^T$
and  ${\bar{\rho}} = [\rho_{11},\rho_{21}]^T$. We then obtain two compact equations for the perturbation in
the phases and densities,
\begin{eqnarray} \label{thetavecdot}
\dot{\bar{\theta}}&=&  -\,\Xi \; \bar{\rho} - \vec v_0  \cdot \nabla \bar{\theta},
\\
\label{rhovecdot}
\dot{\bar{\rho}}&=& \, - \nabla \cdot \left( D \; \nabla \bar{\theta} +  \bar{\rho} \; \vec{v_0} \right) 
- \Lambda \;\bar{\theta} \, ,
\end{eqnarray} 
where the background velocity $\vec{v}_0$ is the same in both condensates. Now combine these two equations into one:
\begin{equation} \label{phaseequation} 
\partial_{t} (\Xi^{-1} \; \dot{\bar{\theta}} ) =
 - \partial_{t} \left(\Xi^{-1} \; \vec v_0 \cdot \nabla \bar{\theta} \right) 
 - \nabla (\vec v_0 \; \Xi^{-1} \; \dot{\bar{\theta}} )  
 + \nabla \cdot \left[ \left(D - \vec v_0 \; \Xi^{-1} \; \vec v_0 \right) \nabla \bar{\theta} \, \right] 
 + \Lambda \; \bar{\theta}.  
\end{equation}
In the next section we show how this equation is analogous to a minimally coupled scalar field in
curved spacetime. 


\section{Emergent spacetime in the hydrodynamic limit}

Instead of keeping the analysis general \cite{2BEC,Ralf}, we now focus on the special case when $\Xi$ is independent of space and time, and on the hydrodynamic limit where $\hat X \rightarrow 0$. Then defining
\begin{equation} \label{tildephase}
\tilde\theta = \Xi^{-1/2}\; \bar\theta,
\end{equation}
equation (\ref{phaseequation}) simplifies to 
\begin{equation} \label{phaseequation2}
\fl 
\partial_{t}^2\tilde{\theta} =
 - \partial_{t} \left(\mathbf{I} \; \vec v_0 \cdot \nabla \tilde{\theta} \right) 
 - \nabla    \cdot   \left(\vec v_0 \; \mathbf{I} \; \dot{\tilde{\theta}} \right)  
 + \nabla \cdot \left[ \left(C_0^2 - \vec v_0 \; \mathbf{I} \; \vec v_0  \right) \nabla \tilde{\theta} \right] 
 + \Omega^2 \; \tilde{\theta},
\end{equation}
where
\begin{equation} \label{Omega2}
 C_{0}^2 = \Xi^{1/2}\; D \;\Xi^{1/2} \, ; 
\qquad \hbox{and} \qquad
\Omega^2 =   \Xi^{1/2} \;\Lambda\; \Xi^{1/2}.
\end{equation} 
Both $C_{0}^2$ and $\Omega^2$ are symmetric matrices. If $ [C_{0}^2, \; \Omega^2] = 0$,
which is equivalent to the matrix equation $ D \; \Xi \; \Lambda = \Lambda \; \Xi \; D$, then they have common eigenvectors.
Assuming (for the time being) simultaneous diagonalizability,  decomposition onto the eigenstates of the system results in a pair of independent Klein--Gordon equations
\begin{equation} \label{KGE}
\frac{1}{\sqrt{-g_{\mathrm{I/II}}}}
\partial_{a} \left\{   \sqrt{-g_{\mathrm{I/II}}} \; (g_{\mathrm{I/II}})^{ab} \; 
\partial_{b} \tilde{\theta}_{\mathrm{I/II}} \right\} + 
\omega_{\mathrm{I/II}}^2 \;
\tilde{\theta}_{\mathrm{I/II}} = 0 \; ,
\end{equation}
where the ``acoustic metrics'' are given by
\begin{equation} \label{metric}
(g_{\mathrm{I/II}})_{ab}=\left( \frac{\rho_{\mathrm{I/II}}}{c_{\mathrm{I/II}}} \right)^{2/(d-1)}
\left[
\begin{array}{ccc}
-\left( c_{\mathrm{I/II}}^2-v_0^2 \right)       &|& -\vec{v_0}^{\,T} \\
\hline
-\vec{v_0}  &|& \mathbf{I}_{d\times d}
\end{array}
\right] \, ,
\end{equation}
and where the overall conformal factor depends on the spatial dimension $d$. 
The metric components depend only on the background velocity $\vec{v}_{0}$, the background densities $\rho_{i0}$,  and the speeds of sound of the two eigenmodes, which are given by
\begin{equation}
\label{e:csq-Xi}
c_{\mathrm{I/II}}^2 = 
\frac{\tr[C_0^2] \pm \sqrt{\tr[C_0^2]^2 - 4 \, \det[C_0^2]}}{2} \,.
\end{equation}
Considering the line element obtained from the acoustic metric (\ref{metric}), it is clear that the speed of sound in the AM takes the role of the speed of light. \\
It is also possible to calculate the eigenfrequencies of the two phonon modes,
\begin{equation}
\omega_{\mathrm{I}}^2 = 0; 
\qquad
\omega_{\mathrm{II}}^2 = \tr[\Omega^2] \, .
\end{equation}
A zero/ non-zero eigenfrequency corresponds to a zero/ non-zero mass for the phonon mode. They both ``experience'' the same spacetime (and so there is a unique metric) if
\begin{equation}
\tr[C_0^2]^2 - 4 \det[C_0^2]=0 \, .
\end{equation} 
The fact that we have an AM representing both massive and massless particles is promising for QGP if we now extend the analysis to high-energy phonon modes so that $\hat X \neq 0$. For the following, we concentrate on flat Minkowski spacetime, by setting $\vec{v}_0 =\vec{0}$.
%

\section{QGP beyond the hydrodynamic limit }
Starting from Eq (\ref{phaseequation}) for a uniform condensate, we set the background velocity to zero, $\vec{v}_0 =\vec{0}$, but keep the quantum pressure term, $\hat X \neq 0$. The equation for the rotated phases 
\begin{equation} \label{tildephase2}
\tilde\theta = \hat \Xi^{-1/2}\; \bar\theta
\end{equation}
 in momentum space is then \cite{QGPAM}
\begin{equation}
\omega^2 {\tilde{\theta}}  = 
\left\{
\sqrt{\Xi+X\; k^2} \;\; [D\; k^2+\Lambda]\;\; \sqrt{\Xi+X\;k^2} 
\right\}\;  \tilde{\theta}  = H(k^2) \; \tilde{\theta} \, .
   \label{eq:new-disp-rel}
\end{equation}
Thus the perturbation spectrum must obey the generalized Fresnel equation:
\begin{equation}
\det\{ \omega^2 \;\mathbf{I} - H(k^2) \} =0 \, .
\end{equation}
That is, the dispersion relations for the phonon modes in a 2-component BEC are
\begin{equation}
\omega_{\mathrm{I/II}}^2 = { \hbox{tr}[H(k^2)] \pm \sqrt{
    \hbox{tr}[H(k^2)]^2 - 4\;  \det[H(k^2)] }\over 2},
\label{eq:tot-disp-rel}
\end{equation}
and a Taylor-series expansion around zero momentum gives
\begin{equation} \label{Taylor} 
\omega_{\mathrm{I/II}}^2 = 
\left. \omega^2_{\mathrm{I/II}} \right\vert_{k \rightarrow 0}  
+ \left. \frac{\d \omega_{\mathrm{I/II}}^2}{\d k^2} \right\vert_{k \rightarrow 0}  \!\! k^2
+ \left. \frac{1}{2} \, \frac{\d^2 \omega_{\mathrm{I/II}}^2}{\d \left(k^2\right)^2} \right\vert_{k \rightarrow 0} \!\! \left(k^2\right)^2
+ \mathcal{O}\left[( k^2)^3 \right] \, .
\end{equation}
These are two dispersion relations in the desired form of Eq (\ref{ext_disp_rel}). Below we will explicitly compute Eq (\ref{Taylor}) up to the fourth order. Given that $\omega_{\mathrm{I/II}}^2 = \omega_{\mathrm{I/II}}^2(k^2)$ only even powers in $k$ are permitted, therefore the dispersion relation is invariant under parity. This is by no means a surprising result, because the GPE (\ref{2GPE}) is also invariant under parity.

We now define the symmetric matrices
\begin{equation} \label{C2}
C^2 = C_0^2 + \Delta C^2\,;  
\qquad \Delta C^2 = X^{1/2}\Lambda X^{1/2}  \, ;
\qquad 
Z^2 =  2 X^{1/2} D X^{1/2} = {\hbar^2\over 2} M^{-2}.
\end{equation}
Note that all the relevant matrices (equations \eref{Omega2} and \eref{C2}) have been carefully symmetrized, and note the important distinction between $C_0^2$ and $C^2$.
Now define
\begin{equation}
c^2 = {1\over2}\tr[C^2] \, ,
\end{equation}
which approaches the speed of sound $c^2 \rightarrow c_0^2$, in the hydrodynamic limit $C^2 \rightarrow C_{0}^2$, (see Eq (\ref{e:csq-Xi})).
The second and fourth order coefficients in the dispersion relations (\ref{Taylor}), for a detailed calculation see~\cite{QGPAM}, are:
\begin{eqnarray} 
 \label{omega_I}
\left.{\d\omega_{\mathrm{I/II}}^2\over\d k^2}\right|_{k\to0} 
&=& c^2 \left[ 1\pm \left\{
 2 \tr[\Omega^2 C_0^2 ] - \tr[\Omega^2]\;\tr[C^2] \over
 \tr[C^2] \tr[\Omega^2]  \right\}
 \right]
= c^2 (1\pm \eta_2) \, ; \\ 
\left.{\d^2\omega_{\mathrm{I/II}}^2\over\d(k^2)^2}\right|_{k\to0} &=&  {\textstyle 1\over \textstyle 2} \Bigg[
\tr[Z^2]  \pm \tr[Z^2] 
\pm 2 \frac{\tr[\Omega^2{C}^2_0]-\tr[\Omega^2]\;\tr[C_0^2]}{\tr[\Omega^2]}\tr[Y^2]
\pm {\tr[C^2]^2- 4\det[C^2_0] \over \tr[\Omega^2] } \nonumber \\
\quad \quad & \mp&
{\tr[C^2]^2\over\tr[\Omega^2]} \eta_2^2
\Bigg] 
= 2 \eta_{4} \left( \frac{\hbar}{M_{\rm eff}} \right)^2 \; ,  \label{omega_II}
\end{eqnarray}
where we set $M_{\rm eff} = \sqrt{m_1 m_2}$.

\section{Lorentz violations from UV physics}

In order to obtain LIV purely due to ultraviolet physics, we demand perfect special relativity for $\hat X \rightarrow 0$. In other words, we require all terms in Eqs (\ref{omega_I}) and (\ref{omega_II}) which would otherwise remain in the hydrodynamic limit to be zero. The constraints we get are:
 \begin{eqnarray}
&&C1:\qquad \tr[C^2_0]^2-4\det[C^2_0]=0 \, ;\\
&&C2:\qquad 2\tr[\Omega^2 {C}^2_0]-\tr[\Omega^2]\tr[C^2_0]=0 \, .
\end{eqnarray}
Beyond the hydrodynamic limit, but imposing $C1$ and $C2$, Eqs (\ref{omega_I}) and (\ref{omega_II}) simplify to:
\begin{eqnarray}
\left.{\d\omega_{\mathrm{I/II}}^2\over\d k^2}\right|_{k\to0} &=& 
{1\over2}\left[\tr[C_0^2] +(1\pm1)\,\tr[\Delta C^2]\right]
= c_0^2 + {1\pm1\over2}\tr[\Delta C^2] \, ,\label{eq:varpi2b}
\end{eqnarray}
and
\begin{equation} \label{eq:varpi4b}
\left.{\d^2\omega_{\mathrm{I/II}}^2\over\d(k^2)^2}\right|_{k\to0} 
=
{\tr[Z^2]  \pm \tr[Z^2] \over 2}\pm\tr[C^2_0]\left(-\tr[Y^2]+
{\tr[\Delta C^2] \over \tr[\Omega^2] }\right) \, .  
\end{equation}
To achieve conditions $C1$ and $C2$ in the 2-component BEC the effective coupling between the hyperfine states has to vanish, $ \tilde{U}_{12}=0$. This can be implemented by imposing a particular transition rate $\lambda = -2 \sqrt{\rho_{10}\;\rho_{20}} \; U_{12}$.
In addition to the fine tuning of $\lambda$, the paramters ($U_{ii},\rho_{i0},m_{i}$) have to be chosen so that the speed of sound simplifies to:
\begin{equation}   \label{eq:c0av} \fl
c_0^2 
={\tilde U_{11} \;\rho_{10}\over m_1} 
=  {\tilde U_{22} \;\rho_{20}\over m_2} 
= \frac{m_2 \rho_{10} U_{11} + m_1 \rho_{20} U_{22} + U_{12} (\rho_{10} m_1 + \rho_{20} m_2) }{2 m_1 m_2 }.
\end{equation}
While one eigenfrequency always remains zero, $\omega_{0,\mathrm{I}}\equiv0$, for the second phonon mode we get
$
\omega_{0,II}^2 = {4 U_{12} (\rho_{10} m_2 + \rho_{20} m_1) c_0^2}/  { \hbar^2}.
$
The mass of the modes are then defined as
\begin{equation} \label{mass_final}
m_{\mathrm{I/II}}^2 = \hbar^2 \omega_{0,\mathrm{I/II}}^2/c_0^4 \,,
\end{equation}
and thus the AM corresponds to one massless particle $m_{\mathrm{I}}=0$ and one massive particle,
\begin{equation} 
m_{II}^2 = {
8 U_{12} (\rho_{10} m_1+\rho_{20} m_2) m_1 m_2 
\over
          [m_2 \rho_{10} U_{11} + m_1 \rho_{20} U_{22} 
         + U_{12} (\rho_{10} m_1 + \rho_{20} m_2)]
} \approx  {m^2  \;
8 U_{12}
\over [U_{11}+2U_{12}+U_{22}]
}
\end{equation}
propagating in the acoustic Minkowski spacetime in the hydrodynamic limit. For higher wave numbers we obtain LIV in the form of Eq (\ref{ext_disp_rel}), and the coefficients $\eta_{2}$ and $\eta_{4}$ for the two modes are: 
\begin{equation} 
\fl 
\eta_{2,\mathrm{I/II}}
={\hbar^2\over4 c^4_0} \; 
{\rho_{10} m_1 + \rho_{20} m_2 \over \rho_{10} m_2 + \rho_{20} m_1} \;
{\omega_{0,\mathrm{I/II}}^2  \over m_1m_2}
\approx\left( {m_{\mathrm{I/II}} \over M_{\rm eff} }\right)^2= 
\left( \frac{ \mathrm{quasiparticle\;mass} } {\mathrm{effective\;Planck\;scale}} \right)^2 \, ;
\label{eta2_final}
\end{equation}
\begin{equation}
\fl
 \eta_{4,\mathrm{I/II}} =  \frac{1}{4}
\left[
\frac{\gamma_{\mathrm{I/II}}\,m_{1}\rho_{10}+\gamma_{\mathrm{I/II}}^{-1}\,m_{2}\rho_{20}}{m_{1}\rho_{20}+m_{2}\rho_{10}}
\right] \, , 
\label{eta4_final}
\end{equation}
where $\gamma_{\mathrm{I}}=1$ and $\gamma_{\mathrm{II}}=m_{1}/m_{2}$ are dimensionless coefficients, and we remind the reader that $M_{\mathrm{eff}}=\sqrt{m_{1} m_{2}}$ is is the scale of Lorentz symmetry breaking. 

From the expression it is clear that the quadratic coefficients (\ref{eta2_final}) are non-universal. While one is always zero, $\eta_{2,\mathrm{I}}\equiv0$, the second $\eta_{2,\mathrm{II}}$ explicitly depends on the interaction constant $U_{12}$. For $U_{12} \ll (U_{11}+U_{22})$, which is equivalent to $m_{\mathrm{II}} \ll \sqrt{m_{1}m_{2}}$, we see that $\eta_{2,\mathrm{II}}$ is suppressed. 
However, there is no further suppression --- after having pulled out a factor $(\hbar / M_{\mathrm{eff}})^2$ --- for the quartic coefficients $\eta_{4,\mathrm{I/II}}$. These coefficients are of order one and non-universal, (though they can be forced to be universal, for example if $\gamma_{\mathrm{I}} = \gamma_{\mathrm{II}}$ and the underlying bosons have equal masses $m_{1}=m_{2}$).  

The suppression of $\eta_2$, combined with the non-suppression of $\eta_4$, is precisely the statement that the ``naturalness problem'' does not arise in the current model.

%
\section{Summary and discussion}
We have presented an AM that can be used as a model for and motivation for conjectures in QGP. Low energy perturbations in a 2-component BEC with laser-induced transitions between the single-particle states reproduce both massive and massless quasi-particles in an emergent spacetime. Furthermore, we have investigated higher energy perturbations for a uniform condensate without background flow. This corresponds to particles propagating in Minkowski flat spacetime with large momentum. While in the hydrodynamic regime the dispersion relation is LI, beyond it the dispersion relation has to be modified in a Lorentz violating manner. 

Due to parity of the GPE we only obtained terms with even exponents of $k$. We calculated the quadratic and quartic dimensionless coefficients $\eta_{2,\mathrm{I/II}}$ and $\eta_{4,\mathrm{I/II}}$. A key observation is that the present model does not suffer from the  naturalness problem,  because the quadratic corrections are Planck suppressed, while at the same time the quartic coefficients $\eta_{4.\mathrm{I/II}}$ have no further suppression, and are actually of order unity.

\section*{Acknowledgements}
This research was supported by the Marsden Fund administered by the Royal Society of New Zealand.


\end{document}